\documentclass[aps,prd,reprint,amsmath,amssymb,superscriptaddress,nofootinbib]{revtex4-2}
\usepackage{graphicx}
\begin{document}
\title{Physics-Constrained Generative Inference of Sub-Crystal
Electromagnetic Shower Structure in a Segmented Calorimeter}
\author{Yu-Sheng Liu}
\affiliation{Department of Physics, National Kaohsiung Normal University,
Kaohsiung 824, Taiwan}
\author{Yu-Chen Tung}
\email[Corresponding author: ]{yctung@mail.nknu.edu.tw}
\affiliation{Department of Physics, National Kaohsiung Normal University,
Kaohsiung 824, Taiwan}
\date{\today}
\begin{abstract}
The finite transverse granularity of a segmented electromagnetic calorimeter fundamentally limits the precision with which the observables of a shower can be reconstructed, among them its position, its lateral profile, and the direction of the incident particle. We show that a substantial fraction of the information suppressed by the segmentation can be inferred under physical constraints, and that it propagates to downstream physics quantities. The reconstruction is cast as an inverse problem and solved with a generative model constrained by the low-order spatial moments of the shower, driving the solution toward physically consistent energy distributions rather than image similarity alone. Using the undoped CsI calorimeter of the KOTO experiment as a reference system, the reconstruction reduces the per-event residual of these moments with respect to the truth-level reference by roughly $40$--$62\%$ in a representative $1$~GeV bin. The inferred morphology also generalizes beyond the training objective: on a $K_{L}\to\pi^{0}\nu\bar\nu$ Monte Carlo sample it improves the reconstruction of the photon incident angle, a quantity never used during training, and of the $\pi^{0}$ decay vertex. These results indicate that finite segmentation is better viewed as a limit on what a calorimeter measures directly than as an absolute limit on the physics information it retains, with the recoverable fraction depending on the observable and growing with the shower energy.
\end{abstract}
\maketitle
\section{Introduction}
\label{sec:introduction}

High-energy physics experiments often rely on electromagnetic calorimeters to measure the kinematic properties of electromagnetically interacting particles, such as photons and electrons. 
While total energy and impact position are the primary observables, the transverse shower profile contains information about shower development that can be exploited to reconstruct higher-level observables.
However, the resolution of these observables is limited by the finite transverse segmentation of the readout: each crystal integrates the deposited energy over its volume and acts as a spatial low-pass filter on the shower. The granularity therefore propagates directly into the reconstructed observables: it limits the precision of the shower position, it broadens the recorded lateral profile, and it suppresses the profile asymmetry that carries information on the incident direction.

The central observation of this work is that the segmentation suppresses this spatial information without entirely eliminating it: because a shower deposits energy in several cells, part of what the segmentation hides remains encoded in the sharing between neighboring channels, and can be inferred when the reconstruction is constrained by the physics of shower development.
Increasing the granularity in hardware, by reducing the cell size or the readout pitch, is mechanically impractical or prohibitively costly in most existing experiments, which motivates computational approaches that reconstruct sub-crystal information and improve the effective granularity of a calorimeter already in operation.
The KOTO experiment provides a concrete example. In the search for the rare decay $K_{L}\to\pi^{0}\nu\bar\nu$~\cite{takuKOTO,ahn2021koto}, the $\pi^{0}$ decay vertex is reconstructed from the two photon energies and their impact positions under the $\pi^{0}$ mass constraint, so the vertex resolution depends on both the photon energy resolution and the shower position resolution. The transverse cell size, determined by cost and fabrication constraints, limits the latter, and thereby propagates into a principal kinematic observable of the experiment. The centroid is one of several observables carried by the low-order spatial moments of the shower, all limited by the same segmentation.

Recovering the sub-crystal energy distribution from the segmented readout is an ill-posed inverse problem: the readout is a coarse observation of an underlying continuous shower. Classical shower-shape parameterizations extract sub-cell information from segmented calorimeters~\cite{fabjan2003,watanabe2005,awes1992}, but rely on fixed analytical models that do not capture the event-by-event fluctuations of the cascade. 
Deep learning has since been applied to calorimeter simulation and reconstruction~\cite{paganini2018,belayneh2020,radovic2018,khattak2022fastsr,krause2021caloflow,mikuni2022caloscore}, and super-resolution methods developed for natural images with convolutional and adversarial models~\cite{dong2016srcnn,goodfellow2014,ledig2017srgan} have been applied to calorimeter showers through flow-based upsampling~\cite{pang2023supercalo} and adversarial super-resolution of photon images~\cite{erdmann2023srgan,arjovsky2017}. These methods optimize a measure of image similarity, in pixel or feature space, which is not equivalent to fidelity of the physics observables: a shower can be close to the reference as an image while still carrying a bias in its total energy, a shift in its centroid, or a distortion of its profile. We instead constrain the low-order spatial moments of the shower directly, so that the residual bias of each observable is a quantity we measure (Sections~\ref{sec:morph_validation} and~\ref{sec:ablation}) rather than one assumed to follow from image similarity.

We develop a physics-constrained super-resolution (SR) framework and demonstrate it on the undoped cesium iodide (CsI) calorimeter of the KOTO experiment at J-PARC. A generative model reconstructs the sub-crystal energy distribution, and the zeroth through third moments of that distribution, the total deposited energy, the centroid, the lateral width, and the profile skewness, enter training as differentiable penalty terms alongside the reconstruction and adversarial losses, restricting the output to physically consistent showers. The reconstruction is not specific to this detector, and the simulated photon energies span the range relevant both to the current KOTO beam ($p_{K_{L}}\simeq 1.4$~GeV/$c$) and to the higher-momentum KOTO~II beam~\cite{koto2_proposal}.

The contribution of this work is threefold. First, the reconstruction is constrained through the low-order spatial moments of the shower, so that physical observables, rather than image similarity, define the objective. Second, using the KOTO CsI calorimeter with its full detector response, the residual bias of each constrained observable is quantified against a truth-level reference. Third, the inferred morphology is shown to generalize beyond the training objective, improving the reconstruction of the photon incident angle, never used in training, and of the $\pi^{0}$ decay vertex on a $K_{L}\to\pi^{0}\nu\bar\nu$ sample.
More broadly, the present study indicates that part of the information suppressed by detector segmentation can be inferred computationally under suitable physical constraints, and propagated to physics observables, extending the reach of an existing calorimeter without hardware modification.

The remainder of this paper is organized as follows. Section~\ref{sec:dataset} describes the calorimeter and the simulated dataset. Section~\ref{sec:methodology} presents the physics-constrained reconstruction framework, including the network, the moment constraints, and the training strategy. Section~\ref{sec:results} presents the physical validation and downstream applications, and Section~\ref{sec:conclusion} concludes.

\section{Calorimeter Geometry and Dataset}
\label{sec:dataset}

\begin{figure}[h] 
   \includegraphics[width=1\linewidth]{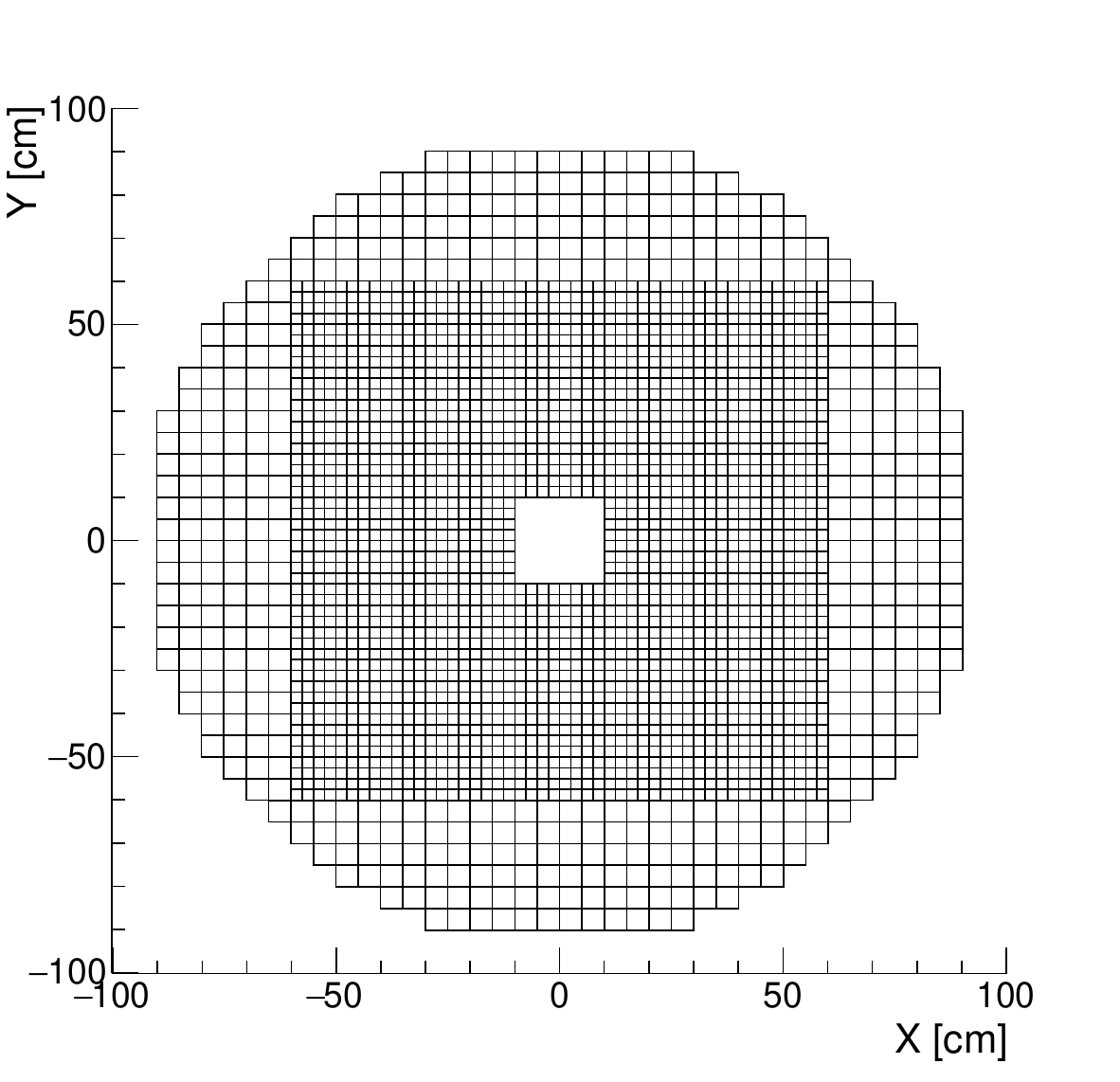}
   \caption{
Schematic layout of the CsI calorimeter viewed from upstream along the beam axis. 
	}
	\label{fig:csi}
\end{figure}

To formalize the physics-constrained SR task, 
we use the undoped CsI calorimeter 
of the KOTO experiment as our experimental reference system~\cite{sato2020csi}.
The calorimeter consists of 2716 CsI crystals arranged in a grid pattern, covering a circular area with a radius of approximately $90$~cm and with a central square beam hole, as shown in Fig.~\ref{fig:csi}. 
Two crystal sizes are employed. 
Crystals of dimensions $2.5\times2.5\times50~\mathrm{cm}^3$ are located in the central $120\times120~\mathrm{cm}^2$ region, while larger crystals of $5.0\times5.0\times50~\mathrm{cm}^3$ are installed in the outer region. 
Each crystal is individually read out, providing a measurement of the energy deposited within its volume. 
The $2.5$~cm transverse cell size sets the scale below which spatial structure is not directly measured, and thereby limits the shower observables that depend on the fine transverse morphology.
The feasibility of reconstructing sub-crystal spatial information stems from the transverse profile of the electromagnetic cascade. The Moli\`{e}re radius of pure CsI ($R_M \approx 3.53$~cm) exceeds the transverse size of the smaller central crystals ($2.5$~cm), so a shower necessarily distributes its energy across multiple adjacent crystals.
The relative energy deposited in each crystal therefore encodes 
the sub-crystal localization of the shower, 
as exploited classically by center-of-energy 
and shower-shape methods~\cite{fabjan2003,watanabe2005,awes1992}. 
The energy sharing among these crystals encodes the sub-crystal position of the shower, and it is this sharing that the reconstruction exploits.

The calorimeter response to a photon is thus a two-dimensional array of crystal energies, which we treat as an image with one pixel per crystal. The finite cell size makes this a coarse-grained sampling of the continuous shower, and the reconstruction task is to infer the finer distribution beneath it. 
The dataset used in this study is generated using the
{\scshape Geant4} simulation toolkit~\cite{agostinelli2003,allison2006,allison2016}, 
where each event corresponds to a single photon originating from the beam axis 
and incident on the calorimeter. 
The photon energy ranges from $50$ to $3000$~MeV, 
with incident angles spanning $1^\circ$ to $80^\circ$. 
Here the incident angle $\theta$ is defined as the polar angle between the photon direction and the front-face normal of the calorimeter, 
with $\theta=0^\circ$ corresponding to normal incidence. 
The azimuthal angle is sampled uniformly.
A total of one million events were simulated and randomly partitioned into $50\%$ training, $25\%$ validation, and $25\%$ test samples.
The simulation incorporates light attenuation and electronics effects, 
producing shower images that 
reflect the realistic detector response of the calorimeter.
Because the calorimeter consists of crystals with different transverse dimensions, 
the resulting images initially contain pixels 
representing both small and large crystals.
To obtain a spatially uniform 
Cartesian grid suitable for convolutional neural networks, 
the large crystals are subdivided into four smaller units 
matching the dimensions of a small crystal, 
with each sub-unit assigned one quarter of the deposited energy, 
assuming a uniform energy distribution within each large crystal volume.
For each event, 
the calorimeter energy response is then cropped to a $21\times21$ pixel image 
centered on the crystal with the maximum deposited energy. 
Each pixel in this array corresponds to the transverse cross-section of a single small crystal, forming the low-resolution (LR) input image.

To establish the targets for supervised learning, high-resolution (HR) images are generated by introducing 
a virtual transverse segmentation within each crystal. 
In this study, each crystal is subdivided into $k \times k$ virtual sub-crystals, where 
$k = 2, 3, 4,$ or $5$, 
corresponding to progressively finer subdivisions 
of the transverse cross-section of a small crystal.
The energy in each virtual sub-crystal is computed directly from the {\scshape Geant4} truth-level energy depositions, prior to the application of light attenuation and electronics effects. The virtual segmentation serves solely as a computational reference for the underlying continuous energy deposition. It does not correspond to any physically realizable detector geometry, and no finer detector is implied.
The resulting HR images constitute the HR reference and therefore serve as physically motivated targets that approximate the underlying shower profile at finer spatial scales, whereas the LR images incorporate the realistic detector response (light attenuation and electronics effects).
This separation defines the reconstruction target of the SR model,
and its implications are discussed in Section~\ref{sec:discussion}.
Supervised training pairs $(I_{\mathrm{LR}}, I_{\mathrm{HR}})$ are constructed for identical simulated shower events, providing a dedicated dataset for studying sub-crystal spatial reconstruction.
The corresponding effective transverse pixel sizes for the LR input and the HR reference 
at each upsampling factor are summarized in Table~\ref{tab:granularity}. 
The SR reconstruction (the model output, described in Section~\ref{sec:methodology}) 
is generated on the same grid as the HR reference
at each $k$ and therefore shares its pixel size.

\begin{table}[h]
\centering
\caption{Effective transverse pixel size of the LR input and of the HR reference (and the SR reconstructions, which are produced on the same grid) for each upsampling factor $k$. The LR pixel corresponds to a single small CsI crystal ($2.5\times2.5~\mathrm{cm}^2$), and each HR level subdivides it into $k\times k$ virtual sub-crystals.}
\label{tab:granularity}
\begin{tabular}{lc}
\hline\noalign{\smallskip}
Level & Pixel size (mm$^2$) \\
\noalign{\smallskip}\hline\noalign{\smallskip}
LR (input)          & $25.0 \times 25.0$ \\
HR, SR ($k=2$)      & $12.5 \times 12.5$ \\
HR, SR ($k=3$)      & $8.33 \times 8.33$ \\
HR, SR ($k=4$)      & $6.25 \times 6.25$ \\
HR, SR ($k=5$)      & $5.0 \times 5.0$ \\
\noalign{\smallskip}\hline
\end{tabular}
\end{table}

\section{Physics-Constrained Reconstruction Framework}
\label{sec:methodology}

\begin{figure*}[t]
	\centering
	\includegraphics[width=1.0\linewidth]{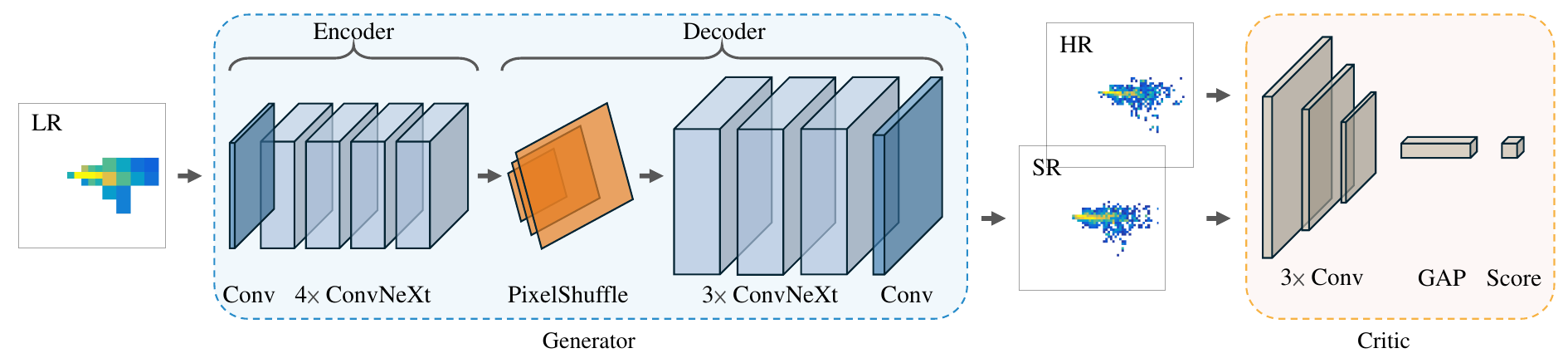}
	\caption{ 	
Architecture of the proposed super-resolution generator and critic. The generator uses ConvNeXt blocks
and PixelShuffle upsampling, while the critic progressively reduces spatial resolution to estimate the Wasserstein critic score of the generated shower.
}
	\label{fig:modelarc}
\end{figure*}

This section describes the framework used to reconstruct the sub-crystal shower distribution from the segmented readout. The reconstruction is first posed as an ill-posed inverse problem (Section~\ref{sec:inverse}) and solved with a generative model whose two components, the generator and the critic, are described in Sections~\ref{sec:generator} and~\ref{sec:critic_adv} and summarized in Fig.~\ref{fig:modelarc}. The physical constraints that act on the generator output, the low-order spatial moments of the shower, are defined in Section~\ref{sec:loss_constraints}, and the training strategy that balances them in Section~\ref{sec:implementation}.

\subsection{The Reconstruction Problem}
\label{sec:inverse}

Recovering the sub-crystal energy distribution from the segmented readout is an inverse problem, and an ill-posed one: the segmentation maps many distinct sub-crystal distributions onto the same coarse image, so the observation does not determine the underlying shower uniquely. Writing the reconstruction as
\begin{equation}
I_{\mathrm{SR}} = \mathcal{G}_w(I_{\mathrm{LR}}),
\end{equation}
where $I_{\mathrm{LR}}$ is the coarse image and $\mathcal{G}_w$ the reconstruction map with parameters $w$, the task is to select, among the distributions consistent with $I_{\mathrm{LR}}$, the one that is physically consistent with an electromagnetic shower.

Two kinds of information make this selection possible. The first is the physics of shower development, which correlates the energy shared among neighboring crystals. 
The second is a set of explicit constraints on the physical observables of the shower, namely the low-order spatial moments of the energy distribution, imposed so that the reconstruction reproduces the quantities actually used in calorimeter analysis. We realize the map $\mathcal{G}_w$ with a generative model trained under these constraints. The role of the generative model is to provide a flexible reconstruction map. The physical content of the method resides in the constraints, and it is these that distinguish the approach from image-based super-resolution. The following subsections describe the two components of the map, the generator and the critic (Fig.~\ref{fig:modelarc}), and then the physical constraints that act on the generator output (Section~\ref{sec:moment}).

\subsection{Generator Architecture}
\label{sec:generator}

The generator is a late-upsampling encoder-decoder built from ConvNeXt blocks~\cite{liu2022convnext}: four blocks extract features in the low-resolution domain, a PixelShuffle layer~\cite{shi2016espcn} performs the spatial upsampling, and three further blocks refine the shower morphology. The large depthwise kernels of the ConvNeXt blocks give a receptive field wide enough to span the transverse spread of an electromagnetic shower, and the fully convolutional design preserves translational equivariance, which suits the stochastic spatial structure of the cascade. From the $21\times21$ low-resolution input the network produces images of $42\times42$ up to $105\times105$ pixels, for upsampling factors $k=2$ to $5$.

\subsection{Critic Architecture and Adversarial Training}
\label{sec:critic_adv}

Adversarial training supplies the stochastic fine structure that deterministic losses alone tend to over-smooth, while the moment constraints keep that structure physically consistent. We use the Wasserstein generative adversarial network (GAN) with gradient penalty (WGAN-GP)~\cite{gulrajani2017}, whose critic estimates a scalar Wasserstein score rather than a classification probability, and which is more stable than the conventional GAN objective for the sparse, spatially concentrated images produced by calorimeter showers. The critic is fully convolutional (three strided convolutional layers followed by global average pooling) and is updated three times per generator step. The corresponding generator term is
\begin{equation}
\label{eq_loss_adv}
L_{\mathrm{adv}} = -\,\mathbb{E}\left[ C(I_{\mathrm{SR}}) \right],
\end{equation}
with $C(\cdot)$ the critic and $\mathbb{E}[\cdot]$ the average over training samples. Because it acts on spatial correlations rather than on individual pixels, the critic drives the generator toward the distribution of the reference.

\subsection{Physics-Constrained Loss Functions}
\label{sec:loss_constraints}

To ensure that the SR model reproduces the physical characteristics of electromagnetic showers, we employ a multi-component composite loss function combining 
two data-driven objectives and four physics-constrained moment terms.
While the two data-driven objectives provide pixel-wise and distributional guidance, 
they do not explicitly enforce the physical regularities of electromagnetic shower development, namely the total deposited energy and the low-order spatial moments of the lateral shower profile. 
These correspond to the zeroth through third moments of the shower energy distribution.
Moments up to third order were selected because they represent physically interpretable calorimeter observables (total energy, centroid, width, and asymmetry) whereas higher-order moments are increasingly sensitive to statistical fluctuations and are rarely used in detector reconstruction. 
The same moment definitions are used for the validation in Section~\ref{sec:morph_validation}.

\subsubsection{Data-Driven Objectives}

\vspace{0.2cm}
\textbf{Generic Reconstruction Loss ($L_{\mathrm{gen}}$):}  
The mean absolute error ($L_1$ loss) is employed as the primary reconstruction constraint. 
Compared to the mean squared error ($L_2$ loss),
the $L_1$ loss is less sensitive to outliers and 
produces less spatial blurring, 
giving more stable guidance for the overall energy distribution of the shower:
\begin{equation}
\label{eq:lgen_def}
L_{\mathrm{gen}} = \frac{1}{N} \sum_{i=1}^{N} \left| I_{\mathrm{SR}}^i - I_{\mathrm{HR}}^i \right| ,
\end{equation}
where $N$ denotes the total number of pixels, and $i$ indexes a pixel location. Equation~(\ref{eq:lgen_def}) states the reconstruction loss in its conceptual form. In the actual implementation this term is restricted to the physically active detector region and is combined with a background-suppression penalty. The corresponding masked expression is given in Section~\ref{sec:bg} (Eq.~(\ref{eq:masked_gen})).

\vspace{0.2cm}
\textbf{Generative Adversarial Loss ($L_\mathrm{adv}$):}  
Pixel-wise losses such as $L_{\mathrm{gen}}$ 
tend to drive the model toward over-smoothed predictions that suppress the stochastic fine-scale structure of the shower. The WGAN adversarial loss,
defined in Eq.~(\ref{eq_loss_adv}) and described in Section~\ref{sec:critic_adv}, is incorporated to overcome this limitation.

\subsubsection{Moment-Based Physical Constraints}
\label{sec:moment}

The four constraints below are the physical content of the method. Each corresponds to a property of the shower that is directly used in calorimeter reconstruction, and together the low-order moments constitute a physically motivated prior: they encode conserved and measurable quantities of an electromagnetic cascade, and require the reconstruction to reproduce them. We state the physical meaning of each before its definition.

\vspace{0.2cm}
\textbf{Zeroth Moment Loss ($L_{\mathrm{M0}}$)} -- Total Deposited Energy:
The total deposited energy is a primary physical observable of the shower. Deviations in total energy between the SR and HR images indicate systematic over- or under-estimation introduced by the generator. This loss penalizes the fractional deviation in total deposited energy:
\begin{equation}
L_{\mathrm{M0}} = \left\langle \frac{\left| \sum_i I_{\mathrm{SR}}^i - \sum_i I_{\mathrm{HR}}^i \right|}{\sum_i I_{\mathrm{HR}}^i} \right\rangle,
\end{equation}
where $I_{\mathrm{SR}}^i$ and $I_{\mathrm{HR}}^i$ denote the pixel values of the SR and HR images, respectively, and $\langle \cdot \rangle$ denotes the average over the training batch.

\vspace{0.2cm}
\textbf{First Moment Loss ($L_{\mathrm{M1}}$)} -- Center of Energy Stability:
The center of energy (CoE) encodes the spatial centroid of the shower and is closely related to the accuracy of downstream kinematic reconstruction. Any artificial spatial shift introduced by the generator propagates as a systematic bias in downstream kinematic observables. This loss penalizes spatial displacement of the CoE between the SR and HR images:
\begin{equation}
L_{\mathrm{M1}} = \left| \bar{x}_{\mathrm{SR}} - \bar{x}_{\mathrm{HR}} \right| + \left| \bar{y}_{\mathrm{SR}} - \bar{y}_{\mathrm{HR}} \right|,
\end{equation}
where the energy-weighted centroid coordinates are defined as:
\begin{equation}
\label{eq:coe}
\bar{x}(I) = \frac{\sum_{i,j} {x_j} \cdot I_{ij}}{\sum_{i,j} I_{ij}}, \quad 
\bar{y}(I) = \frac{\sum_{i,j} {y_i} \cdot I_{ij}}{\sum_{i,j} I_{ij}},
\end{equation}
where $x_j$ and $y_i$ are the physical transverse coordinates of pixel column $j$ and row $i$ (with the pixel pitch given in Table~\ref{tab:granularity}), and $I_{ij}$ is the deposited energy at pixel $(i,j)$. As in $L_{\mathrm{M0}}$, this and the subsequent moment losses are averaged over the training batch, and the batch-average symbol is omitted for brevity.

\vspace{0.2cm}
\textbf{Second Moment Loss ($L_{\mathrm{M2}}$)} -- Lateral Shower Width:
The lateral shower width characterizes the transverse spatial spread of the energy distribution. The coarse granularity of the LR input acts as a spatial low-pass filter that artificially broadens the recorded energy profile, and the SR model must recover the narrower sub-crystal width of the HR reference. This loss penalizes deviations in the shower width along both spatial axes:
\begin{equation}
L_{\mathrm{M2}} = \left| \sigma_{x,\mathrm{SR}} - \sigma_{x,\mathrm{HR}} \right| + 
         \left| \sigma_{y,\mathrm{SR}} - \sigma_{y,\mathrm{HR}} \right|,
\end{equation}
where the lateral shower width is defined as:
\begin{equation}
    \sigma_x(I)
    = \sqrt{
        \frac{\sum_{i,j} \left(x_j - \bar{x} \right)^2 \, I_{ij}}
             {\sum_{i,j} I_{ij}}
    },
\end{equation}
and $\sigma_y(I)$ is defined analogously with $x_j \to y_i$ and $\bar{x} \to \bar{y}$, where $x_j$ and $y_i$ are defined as in Eq.~(\ref{eq:coe}). This quantity governs the lateral extent of the electromagnetic cascade and therefore enters directly into the cluster-shape observables used for particle identification and cluster separation.

\vspace{0.2cm}
\textbf{Third Moment Loss ($L_{\mathrm{M3}}$)} -- Profile Skewness:
The skewness of the shower energy profile reflects the directional asymmetry induced by inclined photon incidence, and is therefore a key observable for angular reconstruction.
Accurately preserving skewness requires the generator to reproduce subtle event-by-event asymmetries in the sub-crystal energy distribution. This loss penalizes deviations in the normalized third central moment:
\begin{equation}
L_{\mathrm{M3}} = \left| \gamma_{x,\mathrm{SR}} - \gamma_{x,\mathrm{HR}} \right| + 
         \left| \gamma_{y,\mathrm{SR}} - \gamma_{y,\mathrm{HR}} \right|,
\end{equation}
where the dimensionless skewness along each axis is defined as:
\begin{equation}
    \gamma_x(I)
    = \frac{\sum_{i,j} \left(x_j - \bar{x} \right)^3 \, I_{ij}}
           {\left( \sum_{i,j} I_{ij} \right)\cdot \sigma_x^3},
\end{equation}
and $\gamma_y(I)$ is defined analogously with
$x_j \to y_i$, $\bar{x} \to \bar{y}$, and $\sigma_x \to \sigma_y$.
The normalization by $\sigma_x^3$ makes $\gamma_x$ independent of the absolute shower width, isolating the directional asymmetry as a purely structural observable. Unlike the lower-order moments, the third moment is particularly sensitive to the asymmetry induced by inclined photon incidence, and it is therefore the observable most closely tied to the directional information examined in Section~\ref{sec:angle}.

\subsubsection{Composite Loss and Validation}
\label{sec:composite}

The generator is trained to minimize the weighted sum of all six loss terms:
\begin{equation}
\label{l_total}
\begin{aligned}
L_{\mathrm{total}} = {}& \lambda_{\mathrm{gen}} L_{\mathrm{gen}} + \lambda_{\mathrm{adv}} L_{\mathrm{adv}} + \lambda_{\mathrm{M0}} L_{\mathrm{M0}} \\
&+ \lambda_{\mathrm{M1}} L_{\mathrm{M1}} + \lambda_{\mathrm{M2}} L_{\mathrm{M2}} + \lambda_{\mathrm{M3}} L_{\mathrm{M3}}.
\end{aligned}
\end{equation}
Balancing these constraints is a non-trivial optimization problem, 
as differences in numerical scale and gradient magnitudes can cause the data-driven terms to dominate the physics-constrained terms. 
The weighting coefficients $\lambda_i$ are fixed once by an empirical calibration protocol that equalizes the weighted contributions $|\lambda_i L_i|$. Because this calibration is tied to the staged optimization schedule, it is described together with the training procedure in Section~\ref{sec:training}, and the resulting values are listed in Table~\ref{tab:loss_weights}.

Monitoring convergence is complicated by the dynamic nature of the adversarial min-max optimization. To address this, the validation loss excludes the stochastic adversarial term and is computed using only the deterministic non-adversarial components:
\begin{equation}
\label{eq:val_loss}
\begin{aligned}
L_{\mathrm{val}} = {}& \lambda_{\mathrm{gen}} L_{\mathrm{gen}} + \lambda_{\mathrm{M0}} L_{\mathrm{M0}} + \lambda_{\mathrm{M1}} L_{\mathrm{M1}} \\
&+ \lambda_{\mathrm{M2}} L_{\mathrm{M2}} + \lambda_{\mathrm{M3}} L_{\mathrm{M3}}.
\end{aligned}
\end{equation}
This physics-based validation strategy provides a stable indicator for model selection and early stopping, preventing adversarial fluctuations from causing premature stopping decisions.

Taken together, the moment terms act as a physical prior on the reconstruction. The inverse problem of Section~\ref{sec:inverse} does not have a unique solution, and the constraints select, among the sub-crystal distributions consistent with the coarse image, those whose low-order moments match the reference. In this sense the total deposited energy, the centroid, the lateral width, and the profile asymmetry play the role that energy conservation or smoothness priors play in other ill-posed reconstruction problems: they encode what is known about electromagnetic showers and restrict the solution accordingly. The generator is thus optimized toward physically meaningful observables rather than toward image fidelity alone.

\subsection{Training Strategy}
\label{sec:implementation}

The framework was implemented in PyTorch~\cite{paszke2019pytorch}. The adversarial optimization and the sparsity of calorimeter images, in which active cells are a small fraction of the total, require care in two respects: the balancing of the competing loss terms, and the suppression of spurious energy in inactive regions. We describe these in turn, after the training schedule. Independent training, validation, and test samples ($50/25/25\%$ of the data) were used throughout.

\subsubsection{Training Schedule and Loss Balancing}
\label{sec:training}

Training proceeds in two stages. The generator is first pre-trained for $10$ epochs on the reconstruction loss alone, so that it learns the global shower morphology before adversarial supervision begins. The full composite loss of Eq.~(\ref{l_total}) is then activated for the remaining training, about $150$ epochs in total.

The balancing of the six loss terms is the step that matters physically, and we fix it by a calibration rather than by hand-tuning. At the start of the second stage all weights are set to unity and the model is run for five calibration epochs. The characteristic magnitude of each term $|L_i|$ is then measured, and the weights $\lambda_i$ are scaled so that the weighted contributions $|\lambda_i L_i|$ are comparable, which prevents any single observable from dominating the generator update. The weights are then frozen, since adjusting them during adversarial training tends to destabilize it. The resulting values are listed in Table~\ref{tab:loss_weights}. Model selection uses the physics-based validation loss of Eq.~(\ref{eq:val_loss}), which combines the reconstruction and moment terms, so that the selected model is the one that best reproduces the physical observables rather than the one with the lowest pixel error.

The optimization uses Adam~\cite{kingma2015adam} ($\beta_1=0.5$, $\beta_2=0.999$), a batch size of $128$, a gradient-penalty coefficient $\lambda_{\rm GP}=10$, and an initial learning rate of $10^{-4}$ reduced by $0.55$ every $30$ epochs, with early stopping (patience $15$) in the final stage. These values are reported for reproducibility, and the results are not sensitive to their precise choice.

\begin{table*}[t]
\centering
\caption{Loss weights $\lambda$ for each upsampling factor $k$.}
\label{tab:loss_weights}
\begin{tabular}{ccccccc}
\hline\noalign{\smallskip}
$k$ & $\lambda_{\mathrm{gen}}$ & $\lambda_{\mathrm{adv}}$ & $\lambda_{\mathrm{M0}}$ & 
$\lambda_{\mathrm{M1}}$ & $\lambda_{\mathrm{M2}}$ & $\lambda_{\mathrm{M3}}$ \\
\noalign{\smallskip}\hline\noalign{\smallskip}
2 & {1.0000} & {0.0219} & {0.8453} & {0.3596} & {0.7133} & {0.0781} \\
3 & {1.0000} & {0.0074} & {0.5031} & {0.0834} & {0.1121} & {0.0250} \\
4 & {1.0000} & {0.0070} & {0.8722} & {0.0631} & {0.1234} & {0.0496} \\
5 & {1.0000} & {0.0058} & {0.9201} & {0.0412} & {0.0782} & {0.0484} \\
\noalign{\smallskip}\hline
\end{tabular}
\end{table*}

\subsubsection{Output Constraints and Background Regularization}
\label{sec:bg}

Electromagnetic shower images are highly sparse, with only a small fraction of pixels containing non-zero energy deposits. Without additional constraints, 
generative models may produce spurious low-energy deposits in inactive detector regions.
A binary activity mask is constructed from the LR input image:
\begin{equation} 
M_{ij} = \begin{cases} 1, & X_{ij} > 0, \\ 0, & X_{ij} = 0 , \end{cases} 
\end{equation}
where $X_{ij}$ denotes the energy deposited in the LR pixel $(i,j)$. 
To be applied to the SR and HR grids, the mask is upsampled by the upsampling factor $k$, so that each active LR cell maps to the corresponding $k\times k$ block of sub-cells.

In practice the pixel-wise reconstruction loss $L_{\mathrm{gen}}$ introduced in Eq.~(\ref{eq:lgen_def}) is evaluated only over the active region defined by this upsampled mask, so that the unrestricted pixel sum of Eq.~(\ref{eq:lgen_def}) is in practice implemented as
\begin{equation}
\label{eq:masked_gen}
L_{\mathrm{gen}} = \frac{1}{\sum_{ij} M_{ij}} \sum_{ij} M_{ij}\left| I_{\mathrm{SR}}^{ij} - I_{\mathrm{HR}}^{ij} \right|,
\end{equation}
which suppresses contributions from physically inactive detector regions. 
In contrast, the moment-based physics constraints (Section~\ref{sec:moment}) are evaluated over the entire SR image, without applying the activity mask, since the spatial moments are global observables that must be integrated over the full energy distribution.

In addition, a rectified linear unit (ReLU) activation is applied at the network output to enforce the non-negativity of deposited energy. To further suppress spurious background activity, an asymmetric background penalty is employed during training. For pixels with zero target energy, the pre-activation prediction $\hat y_{\rm raw}$ is passed through a LeakyReLU function with a negative slope of 0.01 before evaluating the reconstruction error:
\begin{equation} 
\label{eq:lbg}
L_{\rm bg} = \frac{1}{N_{\rm bg}} \sum_{i \in {\rm bg}} \left| {\rm LeakyReLU} \left( \hat y_{{\rm raw},i} \right) \right|, 
\end{equation}
where $N_{\rm bg}$ denotes the number of background pixels with zero target energy. 
Note that the background set in Eq.~(\ref{eq:lbg}) is defined by the HR reference 
(all sub-cells with zero HR-reference energy) and is distinct from the activity mask $M_{ij}$, which is derived from the coarse LR occupancy. An empty sub-cell located inside an active LR cell therefore receives both the masked reconstruction loss and the background penalty.
The corresponding pixel-wise penalty can be written as
\begin{equation} 
|{\rm LeakyReLU}(\hat y_{\rm raw})| 
= \begin{cases} \hat y_{\rm raw}, & \hat y_{\rm raw}>0, \\ -0.01\,\hat y_{\rm raw}, & \hat y_{\rm raw}\le0. \end{cases} 
\end{equation}
The background penalty is treated as part of the generic reconstruction term $L_{\mathrm{gen}}$ in Eq.~(\ref{l_total}) rather than as an independent loss component, 
so the composite objective remains six-component.
This asymmetric constraint intentionally imposes a stronger penalty on positive background artifacts (e.g., non-physical energy deposits), 
which are common in standard, under-constrained GANs. 
Simultaneously, the small non-zero slope ($0.01$) maintains a gradient for negative predictions, preventing gradient saturation, and allowing continued suppression of residual background activity during training.

\section{Physical Validation and Application}
\label{sec:results}

\begin{figure*}[t]
	\centering
   \includegraphics[width=1\linewidth]{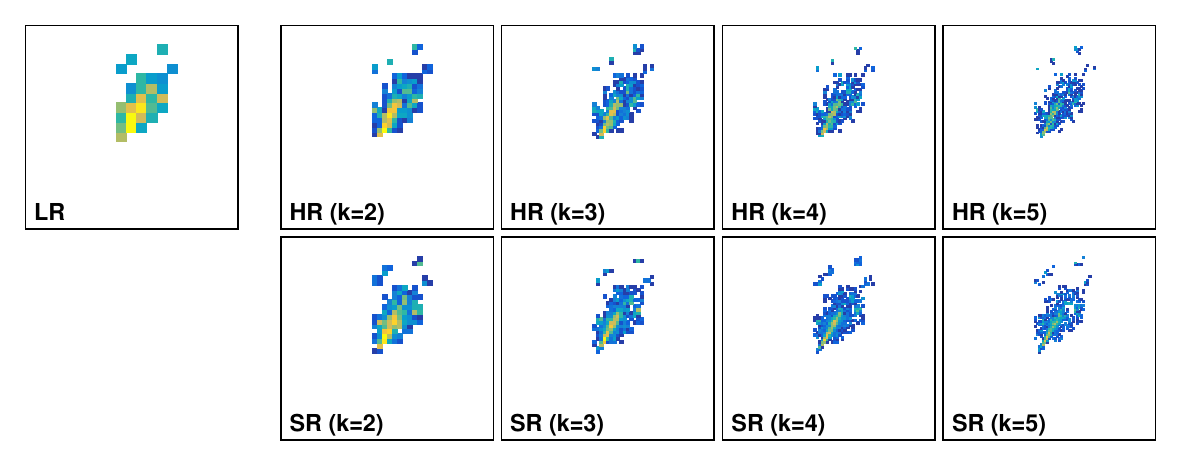}
   \caption{
   \label{fig:csia}
Visual comparison of shower reconstruction for a representative event. The top row shows the LR input followed by the HR reference images at upsampling factors
$k=2$, 3, 4, and 5. The bottom row shows the corresponding SR outputs generated by the proposed model.  
	}
\end{figure*}

\begin{figure*}[t]
    \centering
    \begin{tabular}{@{}c@{\hspace{2mm}}c@{}}
        \includegraphics[page=1,width=0.48\linewidth]{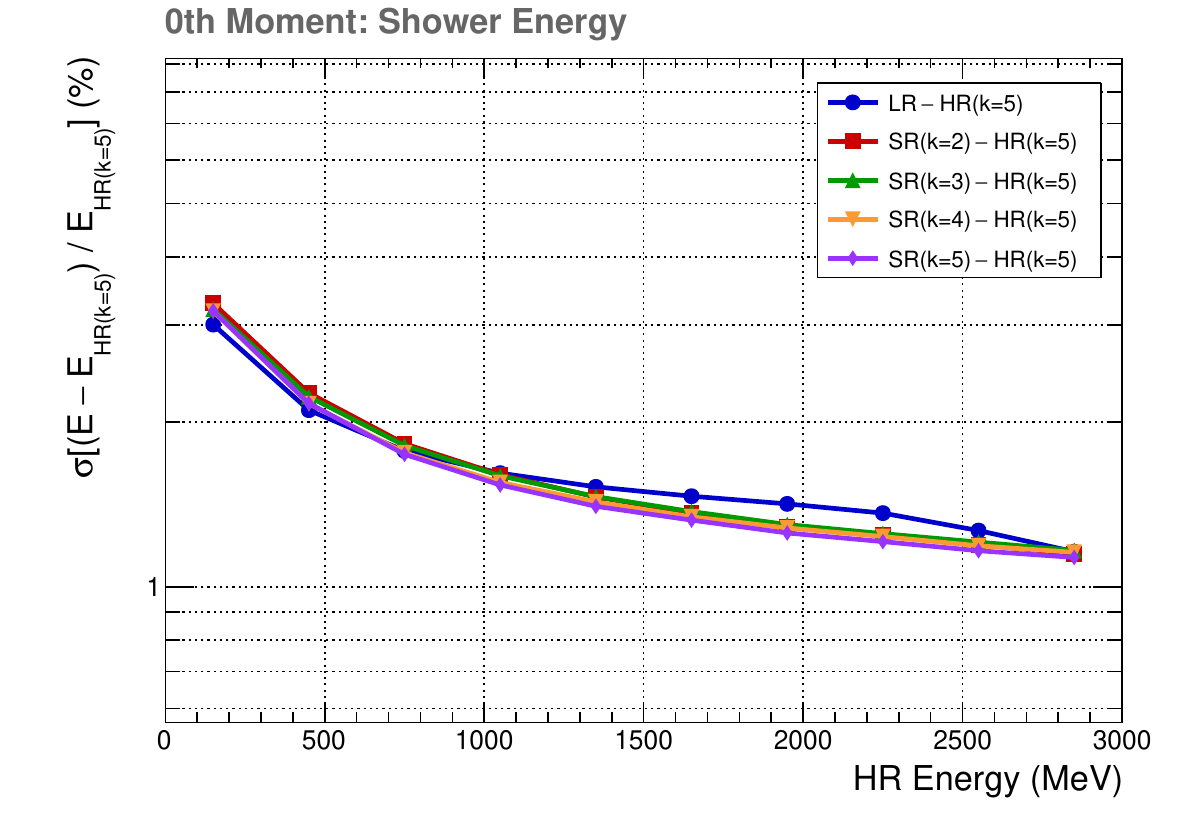} &
        \includegraphics[page=2,width=0.48\linewidth]{./all_plot_v2.pdf} \\     
        \includegraphics[page=3,width=0.48\linewidth]{./all_plot_v2.pdf} &
        \includegraphics[page=4,width=0.48\linewidth]{./all_plot_v2.pdf}
    \end{tabular}

    \caption{
Energy-dependent reconstruction resolution for the four spatial moment observables. 
Each panel shows the 
standard deviation ($\sigma$) of 
the residual distribution, (SR or LR) minus HR($k = 5$), 
as a function of the HR reference energy, 
evaluated in ten energy bins. From top-left to bottom-right: 0th moment (relative total-energy difference $\Delta E/E_{\mathrm{HR}}$), 1st moment (center-of-energy displacement $\Delta \text{CoE}_x$), 2nd moment (shower width $\Delta \sigma_x$), and 3rd moment (skewness $\Delta \gamma_x$).
    }
    \label{fig:cs0i0a}
\end{figure*}

The performance of the proposed physics-constrained SR framework is evaluated both qualitatively and quantitatively. The evaluation proceeds in four steps. First, the SR reconstructions are validated against the HR reference for their fidelity in reproducing the spatial morphology and the low-order moments, the total deposited energy, the centroid, the lateral width, and the skewness (Section~\ref{sec:morph_validation}). Second, an ablation study isolates the contribution of each physics-constrained loss term (Section~\ref{sec:ablation}). Third, the framework is applied to a downstream observable never used in training, the photon incident angle, using an independently trained regression model (Section~\ref{sec:angle}). Fourth, the consequences for the $\pi^{0}$ decay vertex reconstruction are quantified on a $K_{L}\to\pi^{0}\nu\bar\nu$ sample (Section~\ref{sec:zvertex}). Section~\ref{sec:discussion} then discusses the interpretation of the reconstruction target and the principal limitations of the study.

\subsection{Morphological and Spatial Moment Validation}
\label{sec:morph_validation}

We evaluate the reconstruction performance of the SR models across multiple upsampling factors, all trained with the proposed six-component composite loss function. 
The SR reconstructions are evaluated through both visual morphology and the 
spatial-moment framework introduced in Section~\ref{sec:moment}. 
As illustrated in Fig.~\ref{fig:csia},
the generated SR images recover fine-scale spatial structures of the electromagnetic cascade that are absent from the coarse LR inputs. 
The SR framework reconstructs both the dense, sharp shower core and the diffuse peripheral halo, showing good morphological agreement with the HR reference. 
To quantify the reconstruction fidelity of the shower energy distributions across the upsampling factors, 
we compute the moments of order 0 through 3, defined in Section~\ref{sec:moment}, for each shower image and evaluate the residual between the reconstructed SR image and the finest-grid HR($k=5$) reference. 
The finest grid serves as a common reference for all upsampling factors: because the moments are computed in physical coordinates, images of different granularity can be compared directly, although each SR model is trained against the HR target at its own factor $k$.
Because the calorimeter is approximately circularly symmetric and the incident photons are distributed uniformly in azimuth, the shower energy distributions projected 
onto the $x$- and $y$-axes are statistically equivalent. 
The moment-residual analyses in this and the following section are performed 
on the $x$-axis projection. Consistent results are obtained 
for the $y$-axis and are not shown for brevity (the downstream regression in Section~\ref{sec:angle} uses the moments of both projections).

For an energy-dependent resolution analysis, 
the test events are divided into ten bins according to their HR reference energy. 
The reconstruction resolution in each bin is quantified by the standard deviation ($\sigma$) of the per-event residual distribution, providing a fit-independent measure of the spread.
As shown in Fig.~\ref{fig:cs0i0a}, 
five residual curves are presented per plot: 
LR$-$HR($k=5$) (the raw-readout baseline) and four SR($k$)$-$HR($k=5$) curves corresponding to upsampling factors of $k = 2$, 3, 4, and 5.
The SR curves lie at or below the LR baseline across the energy spectrum, with the improvement smallest for the total energy and most pronounced for the higher-order moments.
This trend indicates that the SR reconstruction captures shower features consistent with the HR reference that are not explicitly represented in the original detector segmentation.

The improvement is not uniform across energy. For all three higher moments the SR and LR resolutions both improve with energy, but the SR curves fall faster, so the gap between them widens toward high energy (Fig.~\ref{fig:cs0i0a}). For the centroid, the SR resolution improves from roughly $3$~mm at the lowest energies to below $1$~mm at the highest, while the LR baseline decreases more slowly, so the relative gain grows from about $25\%$ at a few hundred MeV to about $44\%$ near $3$~GeV. The lateral width shows the same pattern most clearly: the LR resolution is nearly flat across the energy range, whereas the SR resolution decreases steadily, so the gain is smallest at low energy and largest at high energy. The skewness follows the same trend.
This common behavior reflects how many cells carry significant signal. At low energy the shower illuminates few crystals, so the transverse morphology is coarsely sampled and little sub-crystal structure remains to be recovered. As the energy rises the shower spreads over more cells, the higher moments become better defined, and the constrained reconstruction has more information to exploit. The total deposited energy is the exception: it is already well measured by the segmented readout at all energies and leaves little to recover, so its residual is essentially unchanged. The size of the gain for each observable is therefore set by how much of it the segmentation removes in the first place. 

\begin{figure}[h]
    \centering
    \begin{tabular}{@{}c@{\hspace{2mm}}c@{}}
        \includegraphics[page=1,width=0.48\linewidth]{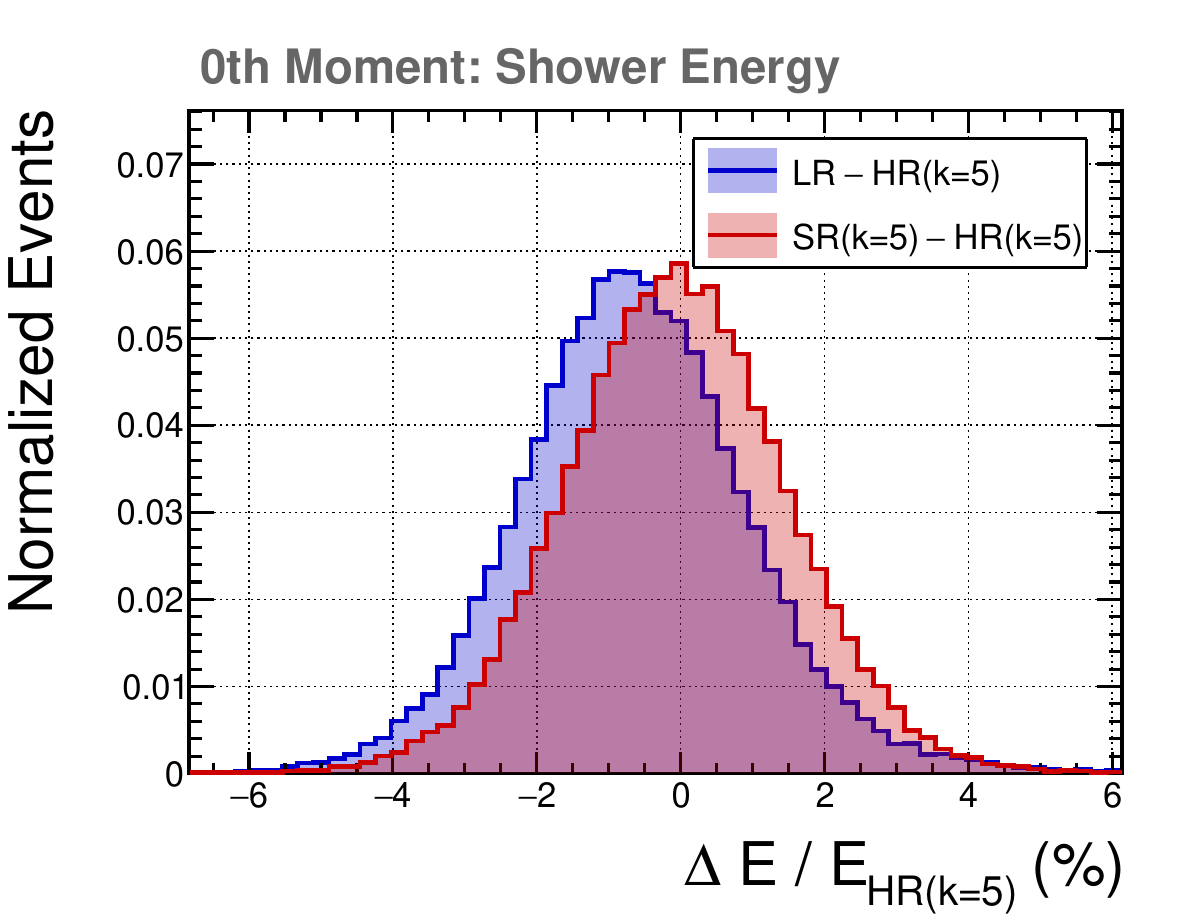} &
        \includegraphics[page=2,width=0.48\linewidth]{./residual_dist_1GeV_v2.pdf} \\     
        \includegraphics[page=3,width=0.48\linewidth]{./residual_dist_1GeV_v2.pdf} &
        \includegraphics[page=4,width=0.48\linewidth]{./residual_dist_1GeV_v2.pdf}
    \end{tabular}
    \caption{Residual distributions for the four moment observables in a representative 1~GeV energy bin, comparing LR$-$HR($k=5$) and SR($k=5$)$-$HR($k=5$).}
    \label{fig:residual_dist}
\end{figure}

To complement the energy-dependent resolution curves, Fig.~\ref{fig:residual_dist} shows the full residual distributions of the four moment observables in a representative $1$~GeV energy bin, comparing the LR input and the SR reconstruction at $k=5$ against the HR reference. The SR residuals are visibly narrower and more symmetric than the LR residuals, confirming that the resolution improvement reflects a genuine sharpening of the per-event observable distributions. For the shower width, the SR reconstruction additionally removes the positive bias that the coarse LR segmentation introduces in the residual.

To verify that the observed improvement does not arise trivially from spatial upsampling alone, we compare the proposed SR framework against bilinear and bicubic interpolation of the LR input in Table~\ref{tab:baseline}. Classical interpolation preserves the centroid but cannot recover the narrower sub-crystal width or the directional skewness, indicating that the physics-constrained SR framework provides information beyond geometric resampling.

\begin{table*}[t]
\centering
\caption{Reconstruction resolution (
$\sigma$, the standard deviation of the per-event residual) for $k=5$ 
in the representative 1~GeV energy bin (the same bin as Fig.~\ref{fig:residual_dist}), compared against classical interpolation baselines. Lower is better. 
$\Delta E/E_{\mathrm{HR}}$ is given in \%, $\Delta\gamma_x$ is dimensionless, and $\Delta\mathrm{CoE}_x$ and $\Delta\sigma_x$ are in mm. 
The bicubic row coincides with the LR row at the quoted precision, as expected for a smooth interpolation that adds no sub-crystal information.}
\label{tab:baseline}
\begin{tabular}{lcccc}
\hline\noalign{\smallskip}
Method & $\Delta E / E_{\mathrm{HR}}~(\%)$ & $\Delta\mathrm{CoE}_x$ & $\Delta\sigma_x$ & $\Delta\gamma_x$ \\
\noalign{\smallskip}\hline\noalign{\smallskip}
LR (raw readout)   & 1.63 & 2.50 & 2.66 & 0.465 \\
Bilinear           & 1.63 & 2.50 & 2.64 & 0.569 \\
Bicubic            & 1.63 & 2.50 & 2.66 & 0.465 \\
Proposed (6-loss)  & 1.56 & 1.42 & 1.01 & 0.261 \\
\noalign{\smallskip}\hline
\end{tabular}
\end{table*}

\subsection{Role of Individual Moment Constraints}
\label{sec:ablation}

To evaluate the physical impact of each constraint term, we perform an ablation study using several reconstruction configurations with progressively increasing levels of physics-constrained supervision. The study focuses on physics observables directly relevant to downstream calorimeter reconstruction, including the total deposited energy, spatial centroid stability, and local shower morphology preservation.
The evaluated configurations consist of: 
(i) a reconstruction-only (1-loss) model trained using solely the generic reconstruction loss ($L_{\mathrm{gen}}$), 
(ii) a baseline adversarial (2-loss) model without explicit physics constraints, optimized via $L_{\mathrm{gen}}+L_{\mathrm{adv}}$, which corresponds to a conventional SRGAN-type baseline~\cite{ledig2017srgan}, and
(iii) the full physics-constrained (6-loss) model defined in Eq.~(\ref{l_total}), 
which additionally incorporates the moment-based physics constraints (M0--M3).

\begin{figure*}[t]
    \centering
    \includegraphics[width=1.0\linewidth, page=1]{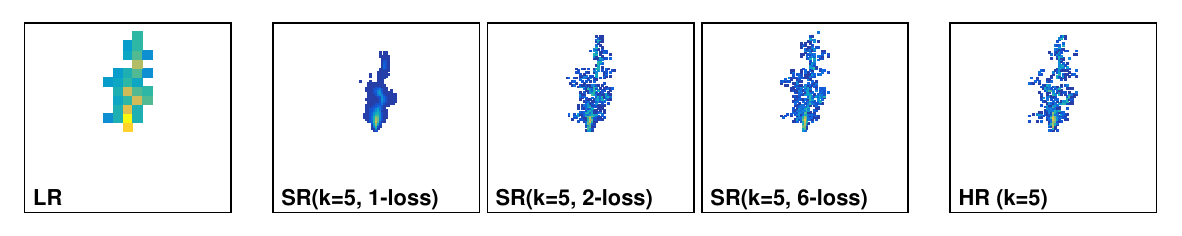} \\
    \caption{
Visual comparison of shower reconstructions for $k = 5$ under different loss configurations. From left to right: LR input, SR($k=5$, 1-loss), SR($k=5$, 2-loss), SR($k=5$, 6-loss), and the HR($k=5$) reference. The loss configurations are compared in the text.
    }
    \label{fig:ablation_visual}
\end{figure*}

The visual reconstructions of these configurations are compared in Fig.~\ref{fig:ablation_visual}. 
While the data-driven 2-loss baseline GAN succeeds in generating fine-scale detail, overcoming the blurriness typical of the pure $L_1$ model, 
these details often manifest as non-physical energy deposits that violate the underlying electromagnetic cascade topology. The addition of the 
moment constraints ($L_{\mathrm{M0}}$--$L_{\mathrm{M3}}$) in the 6-loss configuration stabilizes the four moment distributions.

\begin{figure*}[t]
    \centering
    \begin{tabular}{@{}c@{\hspace{2mm}}c@{}}
        \includegraphics[page=5,width=0.48\linewidth]{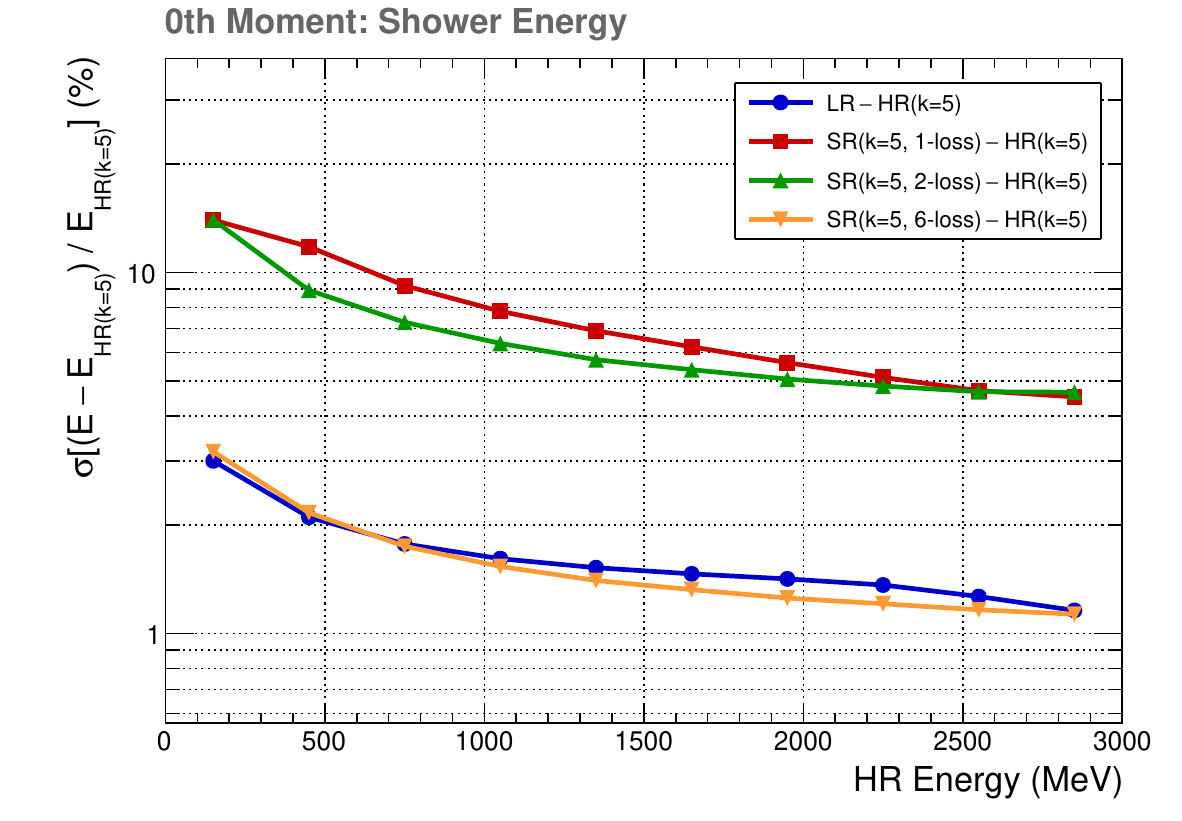} &
        \includegraphics[page=6,width=0.48\linewidth]{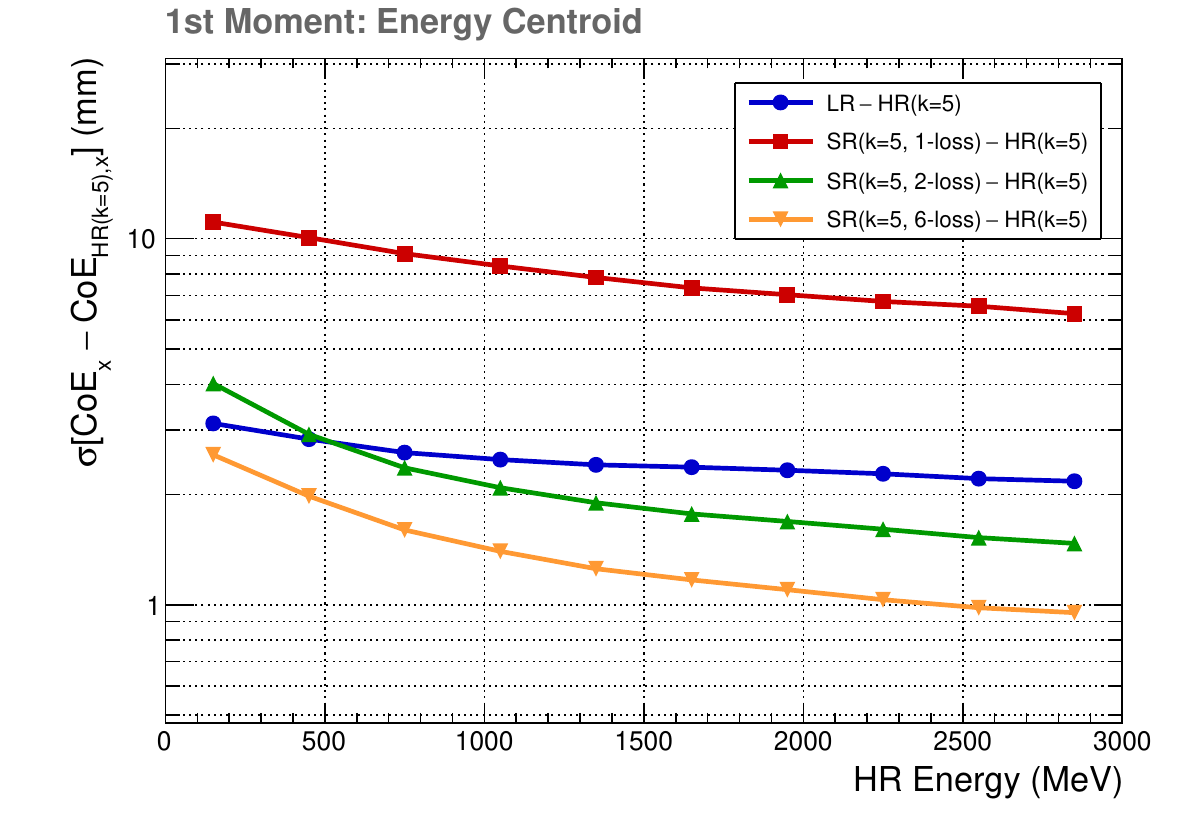} \\
        \includegraphics[page=7,width=0.48\linewidth]{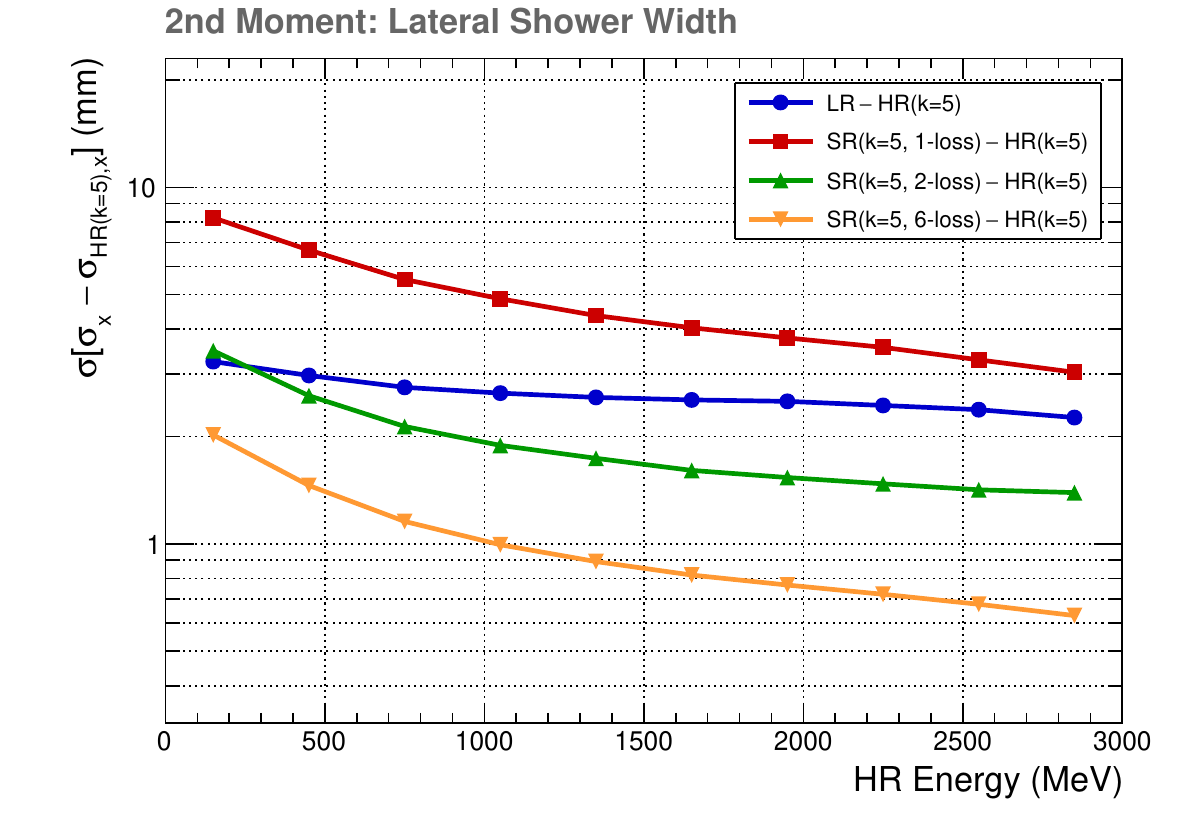} &
        \includegraphics[page=8,width=0.48\linewidth]{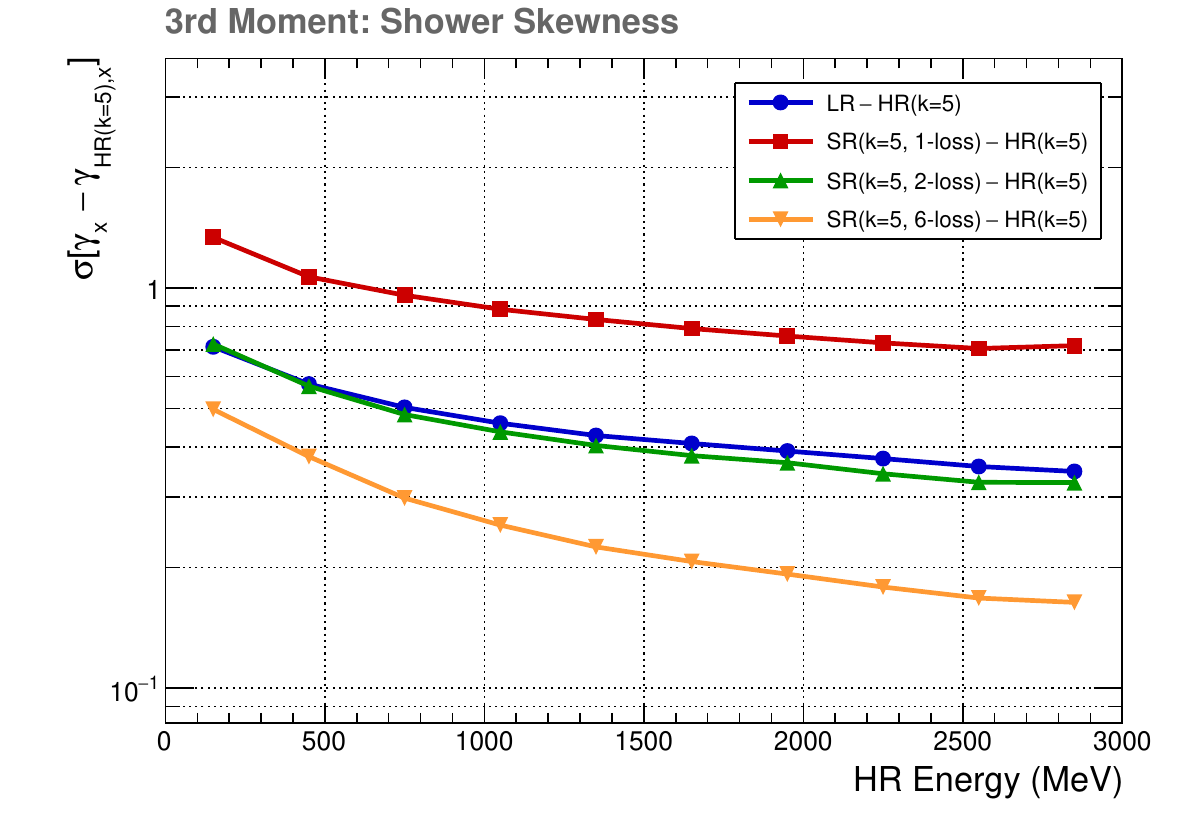}
    \end{tabular}
    \caption{Ablation study of moment reconstruction resolution for $k = 5$. Each panel shows the standard deviation ($\sigma$) of the moment residual, (SR or LR) minus HR($k=5$), as a function of HR reference energy. From top-left to bottom-right: 0th moment 
(relative $\Delta E/E_{\mathrm{HR}}$), 1st moment ($\Delta \text{CoE}_x$), 2nd moment ($\Delta \sigma_x$), and 3rd moment ($\Delta \gamma_x$).}
    \label{fig:ablation_quantitative}
\end{figure*}

The systematic impact of these differences is quantified through the moment residual analysis presented in Fig.~\ref{fig:ablation_quantitative}.
The zeroth-moment (total energy) residual of the 1-loss and 2-loss configurations is substantially degraded relative to the raw LR readout, reflecting the spurious energy deposits introduced in the absence of an explicit energy constraint. 
The $L_{\mathrm{M0}}$ penalty in the 6-loss configuration restores the energy residual to the level reported in Section~\ref{sec:morph_validation} (see also Table~\ref{tab:incremental_ablation}).
The first-moment (CoE) residual also shows improvement in the 6-loss configuration, 
where the explicit $L_{\mathrm{M1}}$ penalty suppresses the centroid shifts introduced by non-physical energy deposits generated by the unconstrained 2-loss baseline. 
By contrast, the second-moment (shower width) and third-moment (skewness) residuals improve markedly only when the full 6-loss objective is applied. The 2-loss baseline exhibits a systematically degraded $\Delta \sigma_x$ residual, driven by the generation of spurious diffuse energy structures, and its skewness $\Delta \gamma_x$ is substantially degraded, performing worse than the LR input itself in some energy regimes.
This degradation in skewness arises because the unconstrained adversarial generator synthesizes spurious asymmetric energy deposits driven by fine-scale spatial fluctuations rather than true directional asymmetry induced by the incident angle. The resulting stochastic pseudo-asymmetry can exceed even the averaging effect of the coarse LR pixels,
underscoring the necessity of explicitly constraining the higher-order shape moments. 
The full 6-loss model reduces these artifacts and achieves moment residuals consistent with the full-model results reported in Section~\ref{sec:morph_validation} across all energy scales. 
The improvement therefore extends beyond pixel-level reconstruction accuracy: without the moment constraints, an adversarial network that produces visually plausible showers can still bias the very observables used in calorimeter measurements, and it is these observables, not image fidelity, that the constraints protect.

The configurations above isolate the combined effect of the moment constraints relative to the unconstrained baselines. To further attribute the improvement to each individual constraint, Table~\ref{tab:incremental_ablation} reports the moment residuals ($k=5$) as the moment terms are added cumulatively from $L_{\mathrm{M0}}$ through $L_{\mathrm{M3}}$. 
Each residual improves when its corresponding constraint is introduced. 
In this energy bin the largest skewness gain arises when $L_{\mathrm{M2}}$ is added, with $L_{\mathrm{M3}}$ providing a further refinement. 
That each constraint predominantly improves its own observable indicates that the moment terms act largely as independent physical constraints on distinct properties of the shower, rather than as a single generic regularizer of the optimization. The one coupling seen here, the improvement of the skewness once the width is constrained, is expected, since the normalized third moment is defined relative to the width. The broader implication is that distinct physical observables require dedicated physical constraints, and are not adequately preserved by a generic image-reconstruction objective.

\begin{table*}[t]
\centering
\caption{Incremental ablation of the moment constraints at $k=5$ in the representative 1~GeV energy bin (the same bin as Fig.~\ref{fig:residual_dist} and Table~\ref{tab:baseline}): moment residual (
$\sigma$) as each physics term is added cumulatively on top of the $L_{\mathrm{gen}}+L_{\mathrm{adv}}$ baseline. $\Delta E / E_{\mathrm{HR}}$ is given in \%, $\Delta\mathrm{CoE}_x$ and $\Delta\sigma_x$ are in mm, and $\Delta\gamma_x$ is dimensionless. Lower is better.}
\label{tab:incremental_ablation}
\begin{tabular}{lcccc}
\hline\noalign{\smallskip}
Configuration & $\Delta E / E_{\mathrm{HR}}$ (\%) & $\Delta\mathrm{CoE}_x$ & $\Delta\sigma_x$ & $\Delta\gamma_x$ \\
\noalign{\smallskip}\hline\noalign{\smallskip}
Baseline ($L_{\mathrm{gen}}+L_{\mathrm{adv}}$) & 6.47 & 2.12 & 1.92 & 0.442 \\
$+\,L_{\mathrm{M0}}$ & 1.57 & 1.91 & 1.36 & 0.382 \\
$+\,L_{\mathrm{M0}}+L_{\mathrm{M1}}$ & 1.56 & 1.43 & 1.20 & 0.360 \\
$+\,L_{\mathrm{M0}}+L_{\mathrm{M1}}+L_{\mathrm{M2}}$ & 1.56 & 1.42 & 1.02 & 0.272 \\
Full ($+\,L_{\mathrm{M3}}$) & 1.56 & 1.42 & 1.01 & 0.261 \\
\noalign{\smallskip}\hline
\end{tabular}
\end{table*}

\subsection{Independent Downstream Validation}
\label{sec:angle}

The preceding sections show that the reconstruction reproduces the constrained moments accurately. A stronger test is whether it recovers a physical observable that was never part of the training objective. We use the photon incident angle for this purpose. A photon leaves no tracking information before it reaches the calorimeter, so its direction can only be inferred from the shower morphology, through the centroid, the lateral spread, and the profile asymmetry. The finite granularity suppresses exactly these shape features, limiting the achievable angular resolution.

The angle is reconstructed from a compact set of physically interpretable inputs: the zeroth through third spatial moments (deposited energy, centroid, lateral width, and skewness) of the reconstructed distribution along both the $x$ and $y$ axes. These are fed to an independently trained XGBoost regressor~\cite{chen2016xgboost}, applied with identical architecture and training to the LR, SR ($k=5$), and HR images, so that any difference in performance originates solely from the quality of the reconstructed observables.

\begin{figure}[!h]
    \centering
        \includegraphics[width=1.\linewidth, page=1]{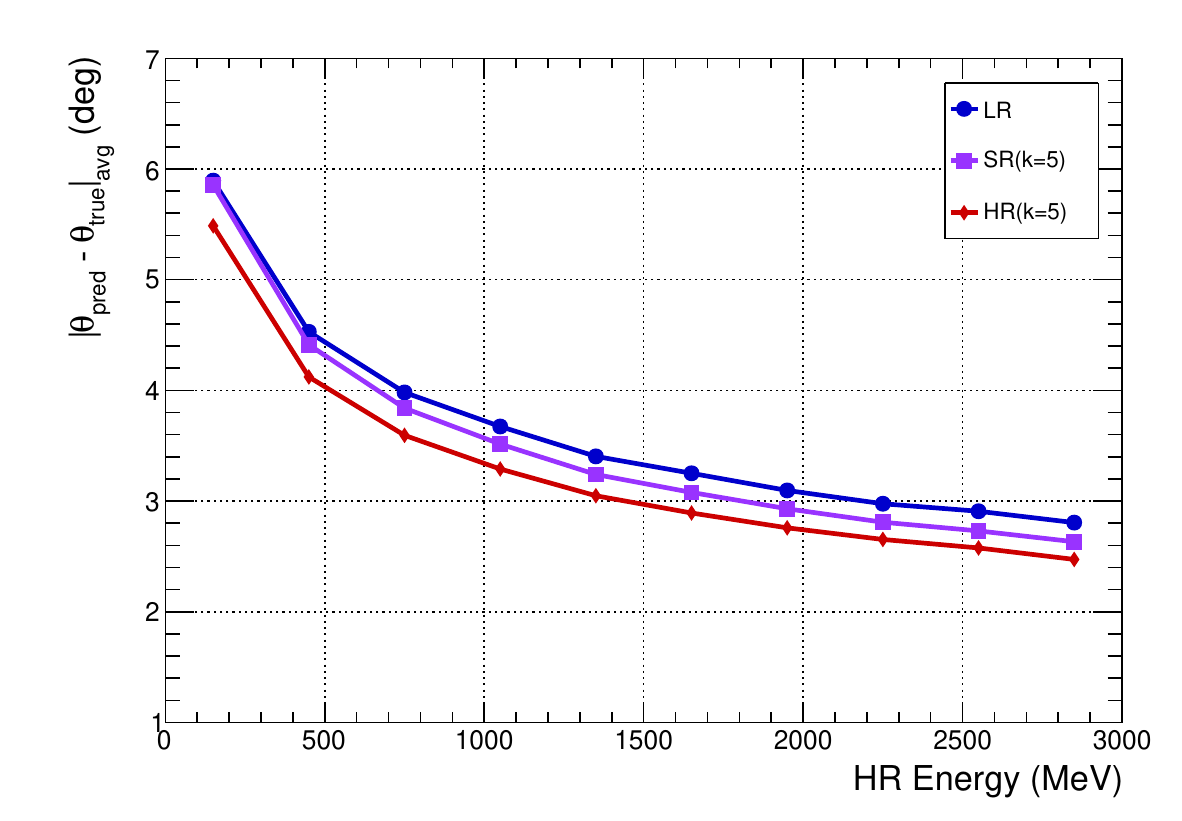}
    \caption{Incident-angle reconstruction performance (MAE) as a function of photon energy, evaluated on a $K_L\to\pi^0\nu\bar\nu$ MC sample with $K_L$ beam momenta representative of the KOTO ($p_{K_L}\simeq 1.4~\mathrm{GeV}/c$) and KOTO II ($p_{K_L}\simeq 3.0~\mathrm{GeV}/c$) configurations. Blue, red, and black points denote the LR, SR ($k=5$), and HR reconstructions, respectively.}
    \label{fig:angle_resolution}
\end{figure}

Whereas the SR model is trained and the morphological validation in Sections~\ref{sec:morph_validation}--\ref{sec:ablation} is performed on a single-photon {\scshape Geant4} sample with the incident angle and energy sampled uniformly, 
the downstream evaluation here is performed on a separate, physics-motivated Monte Carlo (MC) sample that more closely reflects the photon population of an actual rare-kaon-decay experiment. 
In this sample, $K_L$ mesons are generated with momentum distributions spanning the current KOTO beam ($p_{K_L}\simeq 1.4~\mathrm{GeV}/c$)~\cite{ahn2021koto} and the proposed KOTO II beam ($p_{K_L}\simeq 3.0~\mathrm{GeV}/c$)~\cite{koto2_proposal}, and are allowed to decay through $K_L\to\pi^0\nu\bar\nu$. The two photons from the subsequent $\pi^0\to\gamma\gamma$ decay are propagated to the CsI calorimeter, and their truth-level energy and incident angle are recorded. The corresponding LR, SR, and HR shower images are then obtained by applying the SR model trained on the uniform single-photon dataset, so that the trained model is evaluated out-of-distribution with respect to the photon energy and angular spectra of a realistic $K_L\to\pi^0\nu\bar\nu$ event sample. 
The angular reconstruction performance is reported as a function of photon energy in Fig.~\ref{fig:angle_resolution}, quantified by the mean absolute error (MAE), $\langle|\theta_{\mathrm{pred}}-\theta_{\mathrm{true}}|\rangle$. 
The MAE is less sensitive to large-angle outliers, and therefore better reflects the typical (core) reconstruction performance.

Across the photon-energy range covered by the $K_L\to\pi^0\nu\bar\nu$ MC sample, the SR-based reconstruction consistently outperforms the LR baseline and progressively approaches the HR reference. The fraction of the LR--HR performance gap recovered by SR,
\begin{equation}
\label{eq:gap}
f_{\mathrm{gap}} = \frac{\mathrm{MAE}_{\mathrm{LR}} - \mathrm{MAE}_{\mathrm{SR}}}{\mathrm{MAE}_{\mathrm{LR}} - \mathrm{MAE}_{\mathrm{HR}}},
\end{equation}
increases monotonically with photon energy. At photon energies of a few hundred MeV, where the $K_L\to\pi^0\nu\bar\nu$ events from the lower-$p_{K_L}$ beam configuration are concentrated, $f_{\mathrm{gap}}$ remains comparatively modest. As the photon energy grows towards the GeV scale, which is populated primarily by the higher-$p_{K_L}$ beam configuration, $f_{\mathrm{gap}}$ increases steadily and reaches approximately $50\%$ at photon energies above $\sim 2$~GeV. Because the incident angle is never part of the SR objective, this gain is unlikely to be an artifact of the model reproducing a memorized target. It instead indicates that the physics-constrained reconstruction restores directional shower structure that generalizes to an unseen downstream observable and to a photon energy spectrum markedly different from the training distribution. The absolute angular improvement nonetheless remains modest, and smaller in relative terms than the gains on the moments themselves. This is expected: the incident direction is not a constrained quantity but is only indirectly encoded, through the asymmetry of the transverse profile, so the network improves it at one remove from the observables it actually constrains, and only to the extent that the recovered morphology carries directional information. The sub-crystal morphology thus provides directional information analogous to the photon-pointing capability of longitudinally segmented collider calorimeters~\cite{fabjan2003}, obtained here from the transverse profile alone.

\subsection{$K_L \to \pi^0\nu\bar\nu$ Vertex Reconstruction and Beam-Energy Dependence}
\label{sec:zvertex}

The angular reconstruction study of Section~\ref{sec:angle} demonstrates that the SR framework retains directionally relevant shower information on a physics-motivated MC sample. To further quantify the projected physics impact of the method, we evaluate its effect on the longitudinal vertex reconstruction of the $\pi^0$ in $K_L\to\pi^0\nu\bar\nu$ decays, the principal signal channel of the KOTO and KOTO II experiments. The longitudinal decay vertex $Z_{\mathrm{vtx}}$ is a central kinematic observable of the search, and we treat it here purely as a reconstruction observable.

The vertex is reconstructed by the standard method: given the two photon energies and their impact positions on the calorimeter (at $Z_{\mathrm{CsI}} = 6168~\mathrm{mm}$), the $\pi^{0}$ mass constraint fixes the photon opening angle, and the requirement that the two impact positions be consistent with that angle determines the decay-to-calorimeter distance, and hence $Z_{\mathrm{vtx}}$~\cite{ahn2021koto}. The vertex resolution therefore depends on the photon energies and on the two reconstructed positions.

The same $K_L\to\pi^0\nu\bar\nu$ MC sample described in Section~\ref{sec:angle}, spanning the KOTO and KOTO II beam-momentum configurations, is used here. To isolate the impact of photon position resolution on $Z_{\mathrm{vtx}}$ reconstruction, the photon energies entering the vertex reconstruction are taken from the calorimeter-reconstructed values, so that the realistic calorimeter energy resolution $\sigma_E/E$ is fully retained in the evaluation. The photon hit positions are obtained by perturbing the truth-level positions $(x_{\mathrm{true}}, y_{\mathrm{true}})$ with Gaussian smearing of standard deviation $\sigma_{\mathrm{pos}}$. Crucially, the values of $\sigma_{\mathrm{pos}}$ for the LR and SR configurations are not assumed but are extracted directly from the per-photon position residuals measured on the same $K_L\to\pi^0\nu\bar\nu$ MC sample, so that each configuration is evaluated using a position resolution that reflects its actual performance on this physics-motivated event population rather than on the uniform single-photon training sample.

To establish a meaningful reference for the marginal benefit of any further position-resolution improvement, we additionally compute an \emph{energy-only floor}. This floor is obtained by setting $\sigma_{\mathrm{pos}}\to 0$ in the position smearing, so that the residual $Z_{\mathrm{vtx}}$ uncertainty arises solely from the calorimeter energy resolution. It represents an irreducible lower bound on $\sigma_Z$ achievable by improving position resolution alone. Further gains beyond this floor would require improvements in energy resolution. Comparing the LR and SR curves against this floor therefore indicates how much of the current $\sigma_Z$ is driven by position resolution and how much room remains for a position-resolution-based method such as the proposed SR framework.

\begin{figure}[!h]
    \centering
        \includegraphics[width=1.\linewidth, page=1]{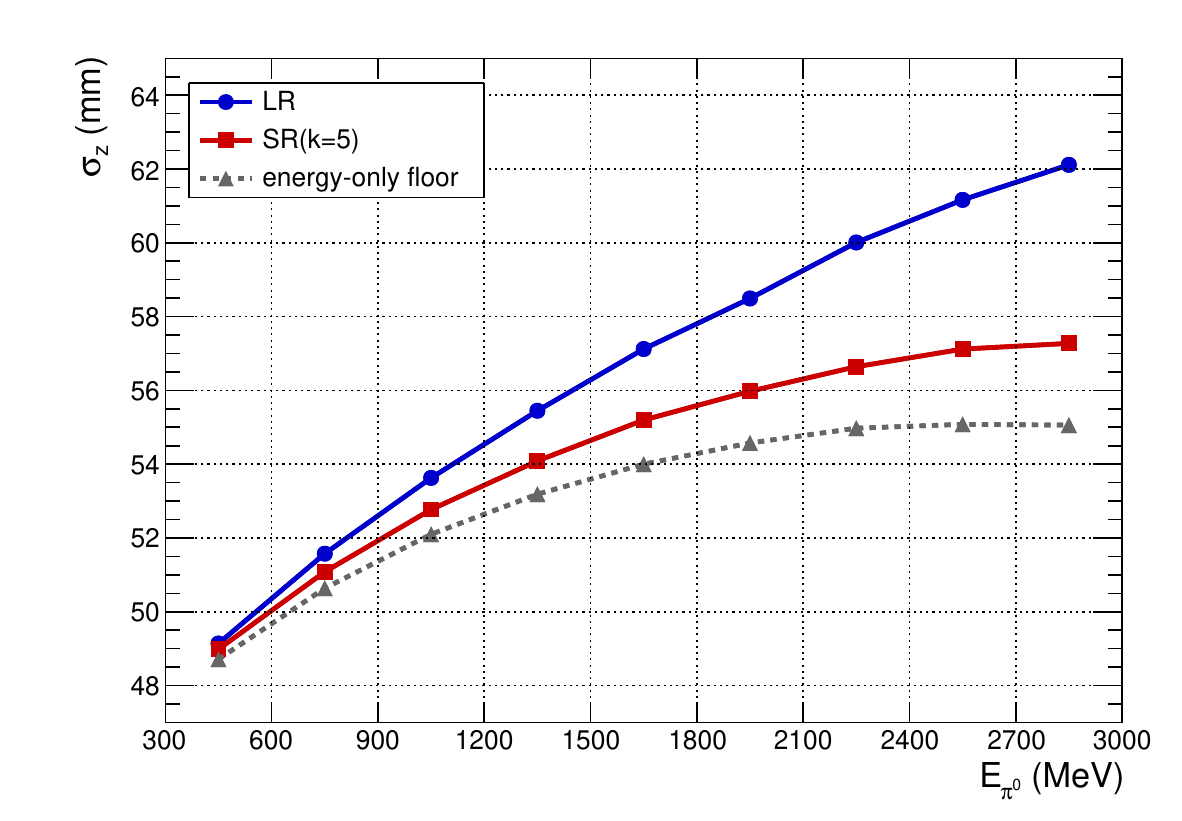}
    \caption{$\pi^0$ vertex resolution $\sigma_Z$ as a function of the reconstructed $\pi^0$ energy $E_{\pi^0}$, evaluated on the $K_L\to\pi^0\nu\bar\nu$ MC sample of Section~\ref{sec:angle}. Blue and red curves denote the LR and SR ($k=5$) configurations and the gray dashed line the energy-only floor ($\sigma_{\mathrm{pos}}\to 0$), all defined in the text.}
    \label{fig:sigma_z}
\end{figure}

Figure~\ref{fig:sigma_z} shows $\sigma_Z$, taken as the width of a Gaussian fitted to the core of the $\Delta Z = Z_{\mathrm{recon}} - Z_{\mathrm{true}}$ residual distribution (iterated over a $\pm 2.5\sigma$ window), as a function of the reconstructed $\pi^0$ energy $E_{\pi^0}$ for the LR configuration, the SR configuration at $k=5$, and the energy-only floor.

Two observations follow from Fig.~\ref{fig:sigma_z}. First, the gap between the LR curve and the energy-only floor widens with increasing $E_{\pi^0}$: at low $\pi^0$ energies the LR position resolution is already close to the floor, leaving little room for any position-resolution-based method to improve $\sigma_Z$, whereas at higher $\pi^0$ energies the LR curve lies substantially above the floor, indicating that position resolution has become the dominant contribution to $\sigma_Z$ and that improvements in position resolution can yield meaningful gains. The origin of this behavior is the two-photon kinematics: the opening angle scales as $m_{\pi^0}/E_{\pi^0}$, so at higher $\pi^0$ energy the two clusters land closer together and a fixed cluster-position resolution maps to a larger relative uncertainty on their separation $d_{12}$, and hence on $Z_{\mathrm{vtx}}$. The high-energy regime in which position resolution limits $\sigma_Z$ is therefore also the regime in which sharpening the reconstructed position is most effective. Second, the SR curve tracks the LR curve closely at low $E_{\pi^0}$ and progressively migrates towards the energy-only floor as $E_{\pi^0}$ increases, recovering an increasing fraction of the LR--floor gap in the high-energy tail.

These two observations delimit the regime in which the method is useful. The SR gain on $\sigma_Z$ is small where the LR curve already lies close to the energy-only floor, at low $E_{\pi^0}$, and grows as the LR curve rises above the floor at higher $E_{\pi^0}$, where position resolution becomes the dominant contribution. The benefit therefore falls on the high-energy part of the $\pi^0$ spectrum, and its per-event size scales with the characteristic $\pi^0$ energy of the host configuration. Since a higher $K_L$ beam momentum shifts the $\pi^0$ spectrum upward, the projected gain increases with beam momentum, and is larger for the KOTO~II configuration ($p_{K_L}\simeq 3.0~\mathrm{GeV}/c$) than for the current KOTO beam ($p_{K_L}\simeq 1.4~\mathrm{GeV}/c$).

\subsection{Discussion}
\label{sec:discussion}

The results admit a clear physical interpretation. Finite detector segmentation does not destroy the sub-crystal information carried by an electromagnetic shower but suppresses it. The energy sharing among neighboring crystals still constrains the underlying shower distribution, while the transverse profile remains governed by well-understood electromagnetic shower physics. The moment constraints encode this physics as a prior, guiding the network toward sub-crystal energy distributions that are statistically consistent with the truth-level reference rather than merely matching it at the pixel level. Because the reconstruction is an ill-posed inverse problem, the inferred distribution should not be interpreted as the unique shower responsible for a given detector response. Instead, it represents one physically plausible realization consistent with both the measured readout and the imposed moment constraints.

Within this framework, the principal finding is that the recoverable information is observable-dependent, and that ordering has a straightforward physical origin. The total deposited energy is already accurately measured by the segmented readout because it is an integral quantity that depends only weakly on how the energy is distributed among neighboring crystals. Consequently, the reconstruction leaves it essentially unchanged. In contrast, the shower centroid and lateral width are directly affected by finite segmentation and recover the largest fraction of the gap to the truth-level reference. The incident direction, although not measured explicitly, is encoded in the asymmetry of the transverse shower profile and is consequently improved through the recovered morphology. More generally, the achievable gain for each observable is determined by how strongly finite segmentation suppresses the relevant information relative to what remains constrained by the detector response. The negligible improvement in the total energy is therefore part of the same physical picture: the method cannot recover information that the detector has already preserved.

The recoverable fraction also increases with photon energy. At low energies, the shower typically extends over only a few crystals, leaving limited sub-crystal structure available for reconstruction. As the energy increases, the shower spreads over more channels, the spatial moments become better defined, and more information survives the segmentation process. The vertex reconstruction study in Section~\ref{sec:zvertex} illustrates the practical consequence of this behavior. The largest improvement is obtained when the low-resolution position resolution remains well above the energy-resolution limit, 
a condition met increasingly toward the high-energy end of the $\pi^0$ spectrum. The potential benefit of the proposed reconstruction therefore increases with the characteristic beam momentum of the experiment.

It is equally important to distinguish the detector limitation addressed here from those that remain fundamentally irrecoverable. Finite segmentation suppresses spatial information while leaving part of it encoded in the energy sharing between neighboring crystals, making partial recovery possible. In contrast, stochastic fluctuations in shower development and scintillation light production are intrinsically random and cannot be recovered by any post-processing algorithm. Likewise, electronics noise, channel-to-channel calibration uncertainties, and cross-talk are absent from the clean simulation used for training and therefore lie outside the scope of the present study. The improvements reported here should consequently be interpreted as the recovery of information suppressed by finite segmentation rather than as an enhancement of the intrinsic calorimeter energy resolution, which the vertex study identifies as the dominant limitation at the low-energy end of the $\pi^0$ spectrum. Furthermore, the reported performance reflects both the recovery of segmentation-suppressed shower morphology and a partial correction of detector-response effects because the truth-level reference is defined before light attenuation and electronics response. The relative contributions of these two effects are not separated in the present work.

The underlying physics is not specific to the KOTO calorimeter. Finite transverse segmentation suppresses shower morphology in any segmented electromagnetic calorimeter, making the same moment-constrained inverse reconstruction applicable to shower-shape and pointing observables in a broad range of existing and future detector systems whose granularity is constrained by cost or engineering considerations, including high-granularity calorimeters. The present study nevertheless remains a simulation-based proof of principle based on single-photon showers and clean $K_L\rightarrow\pi^0\nu\bar{\nu}$ samples. Accidental activity, pileup, calibration uncertainties, and dead channels have not yet been incorporated and should be included before the method is considered for use in an experimental reconstruction chain. Despite these limitations, the present results demonstrate that finite detector segmentation suppresses rather than eliminates part of the information carried by electromagnetic showers, and that a substantial fraction of this information can be recovered through physics-constrained inference.

\section{Conclusion}
\label{sec:conclusion}

The transverse segmentation of an electromagnetic calorimeter suppresses the fine-scale spatial structure of a shower, and with it the precision of the physics observables that depend on that structure. We have shown that a substantial fraction of this suppressed information can be inferred computationally under physical constraints, by constraining the low-order spatial moments of the shower so that physically meaningful observables, rather than image similarity, define the reconstruction objective.

Using the KOTO CsI calorimeter with its full detector response, the per-event residual of these observables with respect to a truth-level reference is reduced, in a representative $1$~GeV bin, by approximately $43\%$, $62\%$, and $44\%$ for the centroid, the lateral width, and the skewness (Table~\ref{tab:baseline}). The inferred morphology also carries information beyond the training objective: the photon incident angle, never used in training, is reconstructed with an accuracy approaching the truth-level reference (Section~\ref{sec:angle}), and the improvement propagates to the $\pi^{0}$ decay vertex, growing with photon energy and largest at the higher beam momentum of the proposed KOTO~II configuration (Section~\ref{sec:zvertex}).

More broadly, the study suggests that finite segmentation is better regarded as a limit on what a calorimeter measures directly than as an absolute limit on the information it retains. The recoverable fraction is set by the physics of the observable and grows with the shower energy, so the benefit is expected to be largest in the high-energy, high-granularity regime of future calorimeters. A demonstration on recorded data, with the accidental activity, pileup, and calibration effects of an operating experiment, is the natural next step.

\begin{acknowledgments}
We thank our colleagues in the KOTO Collaboration for valuable discussions and constructive comments during the development of this work.
\end{acknowledgments}

\section*{Declarations}

\noindent\textbf{Funding.} This work was supported by the National Science and Technology Council (NSTC), Taiwan, under Grant No.~114-2112-M-017-001.

\vspace{4pt}
\noindent\textbf{Competing interests.} The authors declare that they have no competing interests.

\vspace{4pt}
\noindent\textbf{Data availability.} The simulated datasets and the trained models generated in this study are available from the corresponding author on reasonable request.

\vspace{4pt}
\noindent\textbf{Author contributions.} Y.-S.~Liu: Conceptualization, Data curation, Formal analysis, Investigation, Methodology, Software, Validation, Visualization, Writing -- original draft. Y.-C.~Tung: Conceptualization, Formal analysis, Funding acquisition, Investigation, Methodology, Project administration, Resources, Supervision, Validation, Writing -- review \& editing.


\end{document}